# Sex bias in biopsy samples collected from free-ranging dolphins

Sophie Quérouil[1,2,*], Luís Freitas[3], Ana Dinis[3], Filipe Alves[3], Irma Cascão[1], Rui Prieto[1], Mónica A. Silva[1,4], Sara Magalhães[1], José A. Matos[2] and Ricardo S. Santos[1]

[1] *Instituto do Mar (IMAR), Departamento de Oceanografia e Pescas da Universidade dos Açores, Cais Santa Cruz, 9901-862 Horta, Portugal.*

[2] *INETI / Departamento de Biotecnologia, Edifício E - 1º andar, Estrada do Paço do Lumiar 22, 1649-038 Lisboa, Portugal.*

[3] *Museu da Baleia - Município de Machico, Largo Manuel Alves, 9200-032 Caniçal, Madeira, Portugal.*

[4] *4 Biology Department, MS#33, Woods Hole Oceanographic Institution, Woods Hole MA02543, USA.*

\* Corresponding author, present address:

*IRD, UR175 – CAVIAR, Gamet, 361 rue Jean-François Breton, BP5095, F-34196 Montpellier, France. Tel : (+33) 4.67.04.63.36, fax : (+33) 4.67.16.40.04,*
*sophie.querouil@ird.fr*

**Abstract**

Biological samples of free-ranging dolphins are increasingly used to gain information on population structure and ecology. In small cetaceans, the gender of individuals usually cannot be determined at sea, and population sex ratio has to be inferred indirectly. We used molecular sexing to determine the gender of 340 biopsy samples of bottlenose dolphins, *Tursiops truncatus*, spotted dolphins, *Stenella frontalis*, and common dolphins, *Delphinus delphis*, collected around the Azores and Madeira. Sex ratio was globally skewed in favor of males, and differed between species and archipelagos. Skew was probably influenced by the selectivity of biopsy collectors and seasonal or year-round predominance of males in natural populations. Skew was also influenced by sampling duration and intensity. In the Azores, when several samples were successively collected within the same group, the proportion of female samples decreased as a function of sample order. This trend indicated a tendency for females to increasingly avoid the boat while samples were being collected. It showed that males and females reacted differently to the perturbation caused by the biopsy sampling process (*i.e.* sample collection and driving style).

## Introduction

Conservation and management of natural populations require prior knowledge of population structure and dynamics. In Cetaceans, the occurrence of strandings makes it possible to use forensic material to infer population structure. However, such studies are limited to species and places where strandings do occur, and data derived from stranded animals cannot be related to habitat use or social structure (cf. Mesnick et al. 1999). As a consequence, sampling techniques have been developed to obtain samples from free-ranging cetaceans. The most commonly used techniques are skin swabbing (Harlin et al. 1999) and biopsy darting (Krützen et al. 2002). Although more invasive, the latter technique has the advantage of providing a piece of blubber and skin that can also be used to obtain ecological data (*e.g.,* contaminant analyses, biochemical inference of diet).

In dolphins, it is usually not possible to determine the gender of individuals at sea. Population sex ratio has to be inferred indirectly from dead specimens or samples obtained from free-ranging individuals. Stranding and by-catch records revealed a biased sex ratio in several dolphin species and populations (e.g. Perrin and Reilly 1984, Silva and Sequeira 2003; Viricel et al. 2008). However, these records may not accurately represent population sex ratio due to gender segregation in relation to habitat features or differential probability of being caught in fishery nets. Alternatively, gender can be inferred from tissue samples collected from free-ranging dolphins using molecular sexing techniques (*e.g.* Abe et al. 2001). Population genetic studies of dolphins based on biopsy samples do not provide information on sex ratio (most studies) or do not show any significant bias (Krützen et al. 2002, 2004; Möller and Beheregaray 2004; Bilgmann et al. 2007). However, skewed sex ratio was reported in samples of New Zealand's dusky dolphins, *Lagenorhynchus obscurus*, obtained by skin swabbing (Harlin et al. 2003). In that study, sex ratio varied between locations and collecting periods, and was assumed to reflect seasonal variations in space occupation. Another recent study revealed a strongly male-biased sex ratio in samples of bottlenose dolphins, *Tursiops truncatus*, collected by biopsy darting in the Azores (Quérouil et al. 2007). As females with calves were abundant at the time of sampling, the skewed sex ratio was thought not to reflect the actual population structure. It was hypothesized that males were over-represented in the sample due to gender-specific response of dolphins to the sampling process.

In order to determine the frequency and causes of sex bias in biopsy samples collected from free-ranging dolphins, we examined sex ratio variations in 340 biopsy samples of bottlenose dolphins, *Tursiops truncatus*, spotted dolphins, *Stenella frontalis*, and common dolphins, *Delphinus delphis*, collected around the archipelagos of the Azores and Madeira.

## Material and methods

*Sample collection and molecular sexing*

Samples were obtained at two sampling sites in the North-East Atlantic, around the Portuguese archipelagos of the Azores and Madeira. In the Azores, fieldwork was primarily conducted during summer months, between 2002 and 2005. In Madeira, dolphins were sampled all year round, in 2004, 2005 and 2006. A 6-m long inflatable boat was used for most field trips at both locations. In the Azores, 18.9% of samples were collected from a 10-m rigid hull (43 *T. truncatus* and 7 *D. delphis* samples) and in Madeira, 12% were collected from a 19-m sailing boat (2 *T. truncatus*, 9 *S. frontalis* and 4 *D. delphis* samples). Samples were collected using a biopsy darting system (a 125-lb Barnett crossbow, with darts and tips specially designed for small cetaceans by F. Larsen, Ceta-Dart, cf. Mathews et al. 1988), with the exception of one bottlenose dolphin and six spotted dolphin samples that were obtained by the skin swabbing technique. Whenever possible, several individuals from the same group or "school" were sampled (up to 25% of the group count, with a maximum of eight individuals per group). A group was defined as a cluster of conspecific individuals sighted together in a radius of 200 m and involved in similar activities. At sea, groups were sometimes subdivided in sub-groups, with frequent exchanges of individuals between them and/or fission-fusion events. Attempts were made to collect samples from the different sub-groups. However, owing to the instability of sub-group composition, samples were treated at the group-level. In the Azores, mean group size was 22 individuals in *T. truncatus* (+/-s.d.=25), 66 in *S. frontalis* (+/-s.d.=119) and 30 in *D. delphis* (+/-s.d.=39) over the period 1999-2006 (Quérouil et al. 2008). In Madeira, mean group size was 17 individuals in *T. truncatus* (+/-s.d.=17), 18 in *S. frontalis* (+/-s.d.=14) and 17 in *D. delphis* (+/-s.d.=24) over the period 2001-2008 (unpublished data). Sampled groups were significantly smaller than the mean for *S. frontalis* in the Azores (24 +/-s.d.=20; two-sided Student t-test p<0.001) and significantly larger than the mean for *S. frontalis* (48 +/-s.d.=34; p<0.001) and *D. delphis* (43 +/-s.d.=44; p=0.013) in Madeira. Samples were collected from adult-size individuals, but 10 to 30% were probably from immature individuals. In total, 112 samples of bottlenose dolphins, 119 samples of spotted dolphins and 109 samples of common dolphins were analyzed (Table 1).

Samples were processed at the INETI, Lisbon, Portugal. DNA extractions were done following the protocol of Gemmel and Akiyama (1996) or using the DNeasy tissue isolation kit (Qiagen). Molecular sexing was performed by co-amplification of a short fragment of the male-specific SRY gene (CSY, 157 bp) and a microsatellite fragment used as a PCR control for positive identification of females (Abe et al. 2001). The PCR control was Sw15 (234 bp, Richard et al. 1996) for bottlenose dolphins, and GATA028 (Palsbøll et al. 1997) for the other two species. PCR fragments were scanned on an ABI 310 capillary sequencer, using the size marker ROX350 (Applied Biosystems).

*Statistical analyses*

Observed sex ratios were tested against a theoretical 1:1 ratio by means of a Chi-square 2x2 test, separately for each species and archipelago, for each archipelago and for the whole sample. Confidence intervals were calculated based on Wilson's score procedure (Wilson 1927) using the CIProportion.xls spreadsheet available at http://www.cardiff.ac.uk/medic/aboutus/departments/primarycareandpublichealth/ourresearch/resources/index.html (Newcombe 1998). Due to discrepancies between measured sex ratio and field observations, and given that multiple samples were collected within the same groups of dolphins, we hypothesized that the sampling process could influence the sex ratio of the sample. Perturbation was measured in three different ways. 1/ The total duration of the perturbation since group encounter (T-encounter) was defined as the time elapsed between the initial encounter with the group and the collection of a given sample. 2/ The duration of the sampling process (T-sampling) corresponded to the time elapsed between the collection of the first sample obtained within a group and the collection of a given sample (equal to zero for the first sample). Information on T-encounter and T-sampling was available for the Azores only. The effect of species, sex and their interaction on the duration of perturbation prior to sample collection was measured by means of a 2-way ANOVA, after standardization of the dependant variables. Median durations were also compared between male and female samples by means of a Kruskal-Wallis ANOVA by ranks, separately for each species and for the pooled sample. A 2-way ANOVA was also used to test the effect of boat type, sex and their interaction on T-sampling in the Azorean bottlenose dolphins (the only species for which sample size was sufficient to perform the analysis). These tests were performed with Statistica 6.0 (StatSoft). 3/ The intensity of the sampling process was represented by the number of samples already collected at the time a given sample was obtained (*i.e.*, the order in which samples were obtained within a group: first, second, third samples, etc.). As there were only two occasions when more than five samples were collected from the same group (n = 7 and n = 8), the fifth and subsequent samples were pooled. We examined the evolution of the sex ratio of sampled individuals as a function of sample order across all species and sampling locations, and separately for each species and archipelago. We also pooled results by archipelago in order to detect a potential "archipelago effect" and to increase sample size. In each case, we tested for the existence of a correlation between the proportion of females and sample order using Spearman's coefficient (rho). When a significant correlation was found, the data was fitted to a linear or a logarithmic model. The model that provided the best-fit to the data was chosen based on the Akaibe Information Criterion (AIC). To assess how the uncertainty around the proportion of females was affecting the model, the analysis was iterated 1000 times with simulated data. Female proportions were randomly generated in such a way that simulated values followed a normal distribution with 95% of data comprised within the 95% confidence interval defined by Wilson's score procedure. The significance of the regression was calculated for each simulated data set. The percentage of significant regressions was used to estimate the likelihood that the relationship was significant. Computations were performed with R 2.4.1 (R Development Core Team 2006).

**Results and discussion**

*Global sex ratio*

In the whole sample, the overall proportion of females (26.8 %) was significantly lower than 50 % (Table 1). In each species, sex ratio was male-biased in at least one of the two archipelagos. The highest skew (9.1 % females) was found in the common dolphins from Madeira. Skew was moderate and not significant in the common dolphins from the Azores (38.9 % females). A balanced sex ratio (50% females) was found in the bottlenose dolphins from Madeira.

Male-biased sex ratio and seasonal variations in sex ratio are known to occur in natural dolphin populations (*e.g.*, Perrin and Reilly 1984; Silva and Sequeira 2003). However, observations at sea suggested that the observed sex ratios did not adequately reflect the actual population structures. In both archipelagos, numerous females with calves were present in all three species during the sampling season, and populations did not seem to be female-depleted. In the Azores, groups including both adults and calves constituted 64, 76 and 81% of the sightings, respectively for bottlenose, common and spotted dolphins. In Madeira, these groups represented 17, 60 and 52% of the sightings, respectively. Calves could represent as much as 50% of the group count.

It is noteworthy that collectors' selectivity differed between archipelagos. In Madeira, females accompanied by calves were deliberately not sampled, which certainly contributed to limit the overall number of female samples. In the Azores, they were sampled, although often avoided. In this archipelago, and potentially also in Madeira, it can be hypothesized that the low number of female samples was due to boat avoidance by females. In fact, this behavior was quite obvious in spotted dolphins, a species in which segregation by age and sex classes is known to occur (cf. Herzing 1990; Perrin 2002). In both archipelagos, female spotted dolphins accompanied by calves tended to segregate and remain away from the boat, while the rest of the group could be easily approached or deliberately followed the boat. In two instances, the Azorean crew saw groups of spotted dolphins splitting upon arrival of the boat, with females and calves swimming away from it. It is possible that males and immature spotted dolphins come close to the boats in order to divert them from females and their progeny. In the other two species, segregation of females with calves was sometimes observed, but it was not as obvious as in the spotted dolphin.

*Evidence of gender-specific response to the sampling process*

When considering perturbation as the time elapsed since the group was first met (T-encounter, Fig. 1, Azores only), the median duration of perturbation prior to sample collection was significantly influenced by species (2-ways ANOVA: F=36.30, d.f.=2, P<0.0001), but not by sex (F=0.001, d.f.=1, P=0.982) nor by the interaction between species and sex (F=0.81, d.f.=2, P=0.446). Consistently, the K-W ANOVA indicated that T-encounter was not significantly higher for males than for females in all three species (all P>0.05). Inter-specific variations in T-encounter were mainly due to the fact that Azorean bottlenose dolphin samples were collected after an extensive period of photograph collection (typical duration 1-2 hours).

When perturbation was defined as the time since the first sample was collected (T-sampling; Fig. 1, Azores only), the median duration of perturbation prior to sample collection was significantly influenced by sex (2-ways ANOVA: F=4.33, d.f.=2, P=0.039), but not by species (F=2.59, d.f.=1, P=0.078) nor by the interaction between species and sex (F=0.90, d.f.=2, P=0.407). In all species, female samples tended to be obtained more rapidly after the beginning of perturbation than male samples, but this trend was significant only in bottlenose dolphins (z=5.28, P=0.021) and in the pooled sample (z=6.10, P=0.013).

According to the above, the fact that the sampled individual was a male or a female was influenced by the duration of sampling, but not by the total duration since group encounter. This result suggests that collecting samples has a greater impact on dolphins than conducing photo-identification and behavioral studies. This may be caused by the invasiveness of the sampling process itself and/or by a modification in driving style during sampling (e.g. faster speed, more frequent turning and accelerating of the boat, closer distance to the dolphins).

As for the intensity of sampling, the proportion of females decreased as a function of sample order in the pooled sample (Fig. 2). The percentage of females was 33.3% for the first sample collected, and it dropped to about 7% for the fourth and subsequent samples. There was a negative correlation between the two variables (rho = -1; $P$ = 0.017), which was best modeled by a linear regression (% females = -7.3*sample order + 40.8; $P$ = 0.005 and 0.001 for the slope and the intercept, respectively; $R^2$ = 0.945; AIC = 29.3 while the AIC of the logarithmic model was 30.8). Simulations indicated that the uncertainty around the proportion of females did affect the model, as the linear regression was significant in only 44.0% of the simulated data sets.

As a whole, these results suggest that females are more sensitive than males to sampling duration and intensity. A likely explanation of this phenomenon is general avoidance of biopsy sampling boats by females, as opposed to group defense behavior or curiosity of males. This statement is supported by incidental observations. Actually, in the only two instances where more than four samples could be obtained in a group of bottlenose dolphins, sampling was facilitated by an unusual behavior. After the collection of the first biopsies, some dolphins started to closely accompany the boat, and all the biopsies collected from them were males. This is unlikely to result from the fact that large groups are mainly composed of males. Actually, our observations indicated that most groups larger than 20-30 individuals included calves (and thus females).

Differential reaction of males and females to the sampling process had never been reported, but was not unexpected. Previous studies have shown that cetaceans try to avoid interactions with boats when they become too lengthy (Bejder et al. 1999), intrusive (Williams et al. 2002), or unpredictable (Nowacek et al. 2001), and that males and females can display different avoidance strategies (Williams et al. 2002). In bottlenose dolphins, females were shown to adopt a vertical avoidance strategy only when interactions became intrusive and the risk incurred from the interaction was high (Lusseau 2003). Biopsy sampling is a typically intrusive process, and the sex-related response we observed brings additional evidence to the existence of gender-specific avoidance strategies.

*Differences between species and archipelagos*

Although there was a general trend for a decreasing proportion of females along the sampling process, results differed between species and archipelagos (Fig. 3). There was a significant negative correlation between the proportion of females and sample order in the Azores, but not in Madeira. In the Azores, this relationship was best modeled by a linear regression (% females = -10*sample order + 48.5; $P < 0.001$ both for the slope and the intercept; $R^2 = 0.986$; AIC = 25.4 as compared to 35.0 for the logarithmic model). Simulations indicated that the uncertainty around the proportion of females affected the model, as the regression was significant in only 42.5% of the simulated data sets. When species were analyzed separately, a significant negative correlation was obtained only for the Azorean *T. truncatus* samples (Fig. 3). This relationship was best modeled by a logarithmic model (% females = exp(-0.7*sample order + 4.3); $P = 0.015$; AIC = 31.1 as compared to 33.6 for the linear model). Simulations indicated that the uncertainty around the proportion of females greatly affected the model, as the regression was significant in only 9.5% of the simulated data sets.

These results suggest a stronger influence of the sampling process in the Azores than in Madeira. Differences between archipelagos could be due to variations in dolphin behavior and sampling conditions (duration of the sampling process, number of failed attempts, collectors' selectivity, boat and engine type, speed...). It is noteworthy that, in Azorean bottlenose dolphins, the median duration of perturbation prior to sample collection (T-sampling) was not significantly influenced by sex (2-ways ANOVA: F=3.03, d.f.=1, P=0.085), boat (F=2.03, d.f.=1, P=0.158) nor the interaction between boat and sex (F=0.04, d.f.=1, P=0.832). Thus, boat type seems not to have a major influence in this case. Given that sex bias increased along the sampling process in the Azores, we suggest using the sex ratio of the first samples obtained in each group of dolphins as a proxy for population sex ratio. Following this strategy, we propose the following estimates of the proportion of females in each population: 36.4 % for *T. truncatus*, 32.4 % for *S. frontalis* and 48.1 % for *D. delphis* in the Azores, and 50 % for *T. truncatus*, 22.7 % for *S. frontalis* and 9.1 % for *D. delphis* in Madeira (Table 1). Proportions are certainly underestimated in Madeira, because females with calves were not sampled in this archipelago. They are potentially also underestimated in the Azores where females with calves were sampled only occasionally. In the case of the bottlenose dolphins from the Azores, the proportion of females might have been lowered because sampled individuals were chosen among dolphins bearing distinctive dorsal fins, to allow individual identification. It is noteworthy that males bear more marks than females (Smolker et al. 1992), hence have a higher probability of being sampled. The fact that Azorean bottlenose dolphin samples were collected after an extensive period of photograph collection did not have a major influence, as there was no significant difference in the median duration since encounter between male and female samples (Fig. 1). Interestingly, for those bottlenose dolphins for which an age class could be estimated, the percentage of females was 16.1% among adult samples (95% IC = 7.1-32.6%) and 40.9% among subadult samples (95% IC = 23.2-61.3%), confirming that adult females were less likely to be sampled.

The proportion of females varied substantially between archipelagos in common dolphins. The estimated proportion of females was very low in Madeira (9.1 %). These samples differed from the others in that they were collected during winter and spring (Table 1). Interestingly, this is the period when the highest male bias was observed in common dolphins stranded along the continental Portuguese coast (Silva and Sequeira 2003). Conversely, a genetic analyses of common dolphins that died during a massive stranding in February 2002 in the English Channel revealed that all adult individuals were females (N = 47; Viricel et al. 2008). Altogether, these results indicate that segregation by age and sex occurs in common dolphins, and depends on season and location. However, unpublished data from Madeira Whale Museum indicated a balanced sex ratio (1:1) in 28 common dolphins stranded along the coast of Madeira over the period 1995-2006 (95% IC = 32.6-67.4% females; almost all strandings occurred during the winter season). Further work is required to determine the spatio-temporal pattern and environmental determinants of age and sex segregation in *D. delphis*.

*Main findings*

Sex ratio was strongly biased in favor of males in 340 samples collected from free-ranging dolphins in the Azores and Madeira. Bias was shown to increase with the duration and intensity of the sampling process in the Azores. Thus, we recommend that the duration of sampling and the number of samples collected within a group shall be minimized in order to limit perturbation. The amount of bias and the influence of the sampling process varied between species and archipelagos. While bias could reflect seasonal or year-round predominance of males in some populations, it was also influenced by other factors, such as selectivity of the biopsy collectors and dolphin behavior (avoidance of boats by females, curiosity and/or defense behavior of males).

Results suggest that biopsy sampling is not a good method to estimate sex ratio in natural populations. If biopsy samples are to be used to estimate population sex ratio, it appears necessary to search and correct for potential biases, for instance by considering only the first sample obtained in each sampling bout. It is unclear whether less invasive sampling methods, such as skin swabbing (Harlin et al. 1999) or use of a biopsy pole (Bilgmann et al. 2007), provide better estimates of population sex ratio. Their limited range only allows sampling individuals that are at short distance from the boat. In the present study, all six spotted dolphin samples obtained by skin swabbing were males, while the single bottlenose dolphin sample was from a juvenile female. By contrast, no sex bias was revealed in 76 *D. delphis* (Bilgmann et al. 2007), 33 *T. aduncus* (Möller and Beheregaray 2001) and 28 *Tursiops* sp. (Möller et al. 2008) samples collected with a pole in Australian coastal waters. Owing to variations in dolphin behaviors among species and areas, we recommend that each situation shall be evaluated individually.

**Acknowledgements**

Authors wish to thank D. Lusseau, E. Baras and three anonymous referees for useful comments on a previous version of this manuscript. They acknowledge the help of E. Paradis for statistical analyses with R. They are very grateful to the Portuguese Foundation for Science and Technology (FCT) and the FEDER program for funding the CETAMARH (POCTI/BSE/38991/01) and the GOLFINICHO (POCI/BIA-BDE/61009/2004) projects, S.Q.'s post-doctoral grants (IMAR/FCT- PDOC-006/2001-MoleGen and SFRH/BPD/19680/2004), M.A.S.'s doctoral (SFRH/BD/8609/2002) and post-doctoral (SFRH/BPD/29841/2006) grants, S.M.'s investigation assistant grant (CETAMARHII/POCTI/BSE/38991/2001) and I.C.'s investigation assistant grants (IMAR/FCT/GOLFINICHO/001/2005 and IMAR/FCT/GOLFINICHO/004/2006). They also acknowledge FCT for its pluri-annual funding to Research Unit #531 and the EU funded program Interreg IIIb for funding the MACETUS project (MAC/4.2/M10) as well as R.P. and S.M.'s grants (IMAR/INTERREGIIIb/MACETUS/MAC1/2). They wish to thank all the students and staff who contributed to the three projects, with special thanks to the skippers (P. Martins, V. Rosa, R. Bettencourt, N. Serpa, H. Vieira and J. Viveiros), whose dexterity greatly facilitated sample collection.

**Ethical standards**

Sample collection and analysis complied with the current Portuguese laws. Samples were obtained under sampling permits 06/CN/2002, 11/CN/2003, 3/CN/2004 and 7/CN/2005 of the Environment Directorate of the Regional Government of the Azores, and Of. 668/04 Inf 711/04 DAC/DSCN, Credential nº 103-107/2006/CAPT from the Instituto de Conservação da Natureza.

**Conflict of interest**

The authors declare that they have no conflict of interest with the institutions that support their research.

**Table 1.** Number of samples (N) collected by species and archipelago, main sampling period (>2 samples/month), proportion of females with 95% confidence interval (CI), and Chi-2 test of deviation against a 1:1 sex ratio, for all the samples (S = All) and the first samples obtained in each group pf dolphins (S = 1$^{st}$). Best estimates of population sex ratio are in bold. Significant *P*-values are marked with asterisks (*: *P*<0.05, **: *P*<0.01, ***: *P*<0.001).

| Species | Archipelago | Sampling period | S | N | % females | 95% CI | Chi-2 | P |
|---|---|---|---|---|---|---|---|---|
| *T. truncatus* | Azores | May-October | All | 86 | 25.6 | 17.5-35.7 | 20.5 | <0.001*** |
| | | | 1$^{st}$ | 44 | **36.4** | **23.8-51.1** | 3.3 | 0.070 |
| | Madeira | April-September | All | 26 | **50.0** | **32.1-67.9** | 0.0 | 1.000 |
| | | | 1$^{st}$ | 16 | 50.0 | 28.0-72.0 | 0.0 | 1.000 |
| *S. frontalis* | Azores | July-September | All | 75 | 26.7 | 18.0-37.6 | 16.3 | <0.001*** |
| | | | 1$^{st}$ | 37 | **32.4** | **19.6-48.5** | 4.6 | **0.033*** |
| | Madeira | June-November | All | 44 | **22.7** | **12.8-37.0** | 13.1 | **<0.001*** |
| | | | 1$^{st}$ | 34 | 25.0 | 12.0-44.9 | 6.0 | 0.014* |
| *D. delphis* | Azores | July-September | All | 54 | 38.9 | 27.0-57.2 | 2.7 | 0.104 |
| | | | 1$^{st}$ | 27 | **48.1** | **30.7-66.0** | 0.1 | 0.847 |
| | Madeira | January-June | All | 55 | **9.1** | **4.0-19.6** | 38.8 | **<0.001*** |
| | | | 1$^{st}$ | 21 | 4.8 | 0.8-22.7 | 17.2 | <0.001*** |
| All | Azores | | All | 215 | 29.3 | 23.6-35.7 | 36.8 | <0.001*** |
| | | | 1$^{st}$ | 108 | 38.0 | 29.4-47.4 | 6.3 | 0.012* |
| | Madeira | | All | 125 | 22.4 | 16.0-30.5 | 38.1 | <0.001*** |
| | | | 1$^{st}$ | 61 | 24.6 | 15.5-36.7 | 15.7 | <0.001*** |
| All | both | | All | 340 | 26.8 | 22.3-31.7 | 73.4 | <0.001*** |
| | | | 1$^{st}$ | 169 | 33.1 | 26.5-40.5 | 19.2 | <0.001*** |

**Figure 1.** Mean duration of perturbation before collection of male and female samples in the Azores, with standard deviation (bars). A. T-encounter: time since first encounter with the group; B. T-sampling: time since collection of first sample.

**A.**

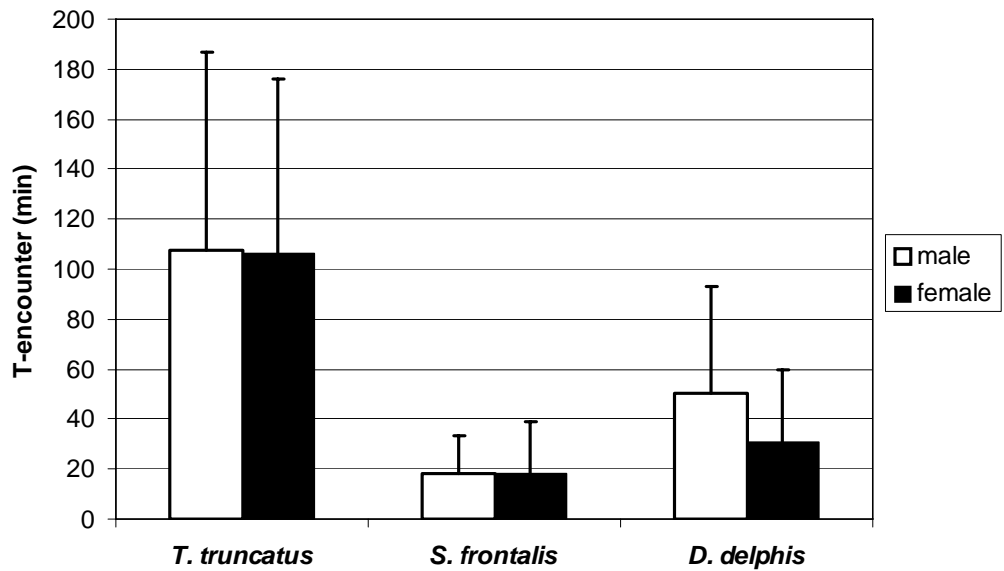

**B.**

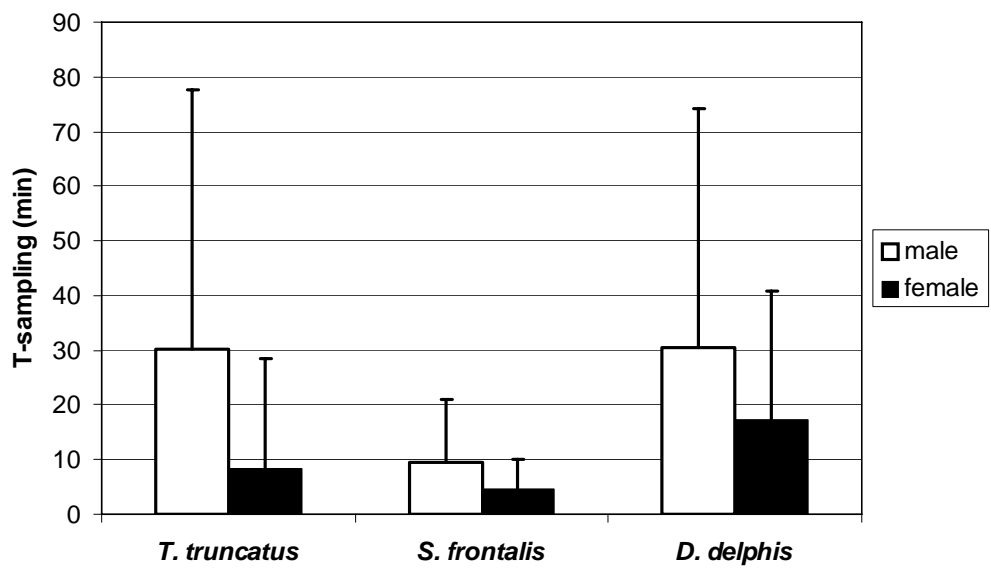

**Figure 2.** Percentage of females as a function of sample order among samples successively collected within groups, with 95% CI (error bars) and fit of a linear regression (dashed line).

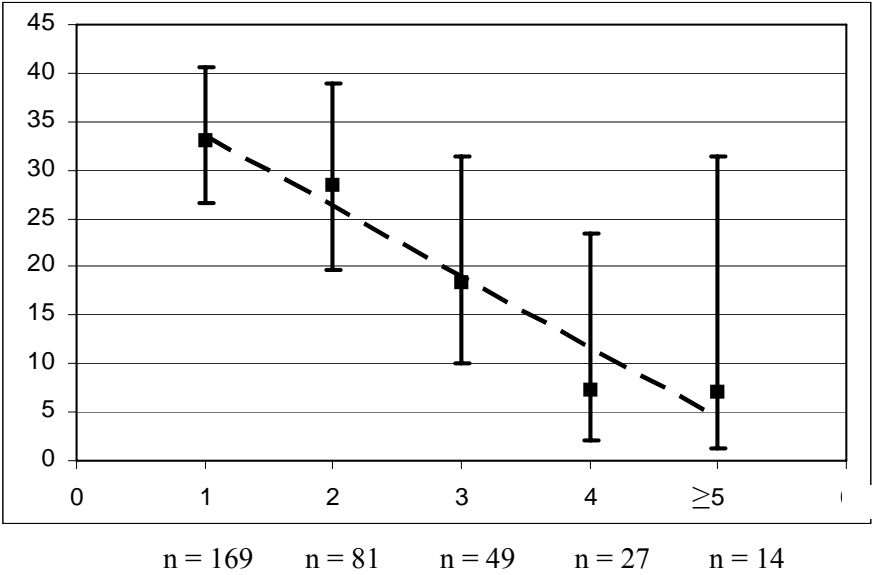

n = 169   n = 81   n = 49   n = 27   n = 14

**Figure 3.** Number of male (open) and female (filled) samples as a function of sample order in each species and archipelago.

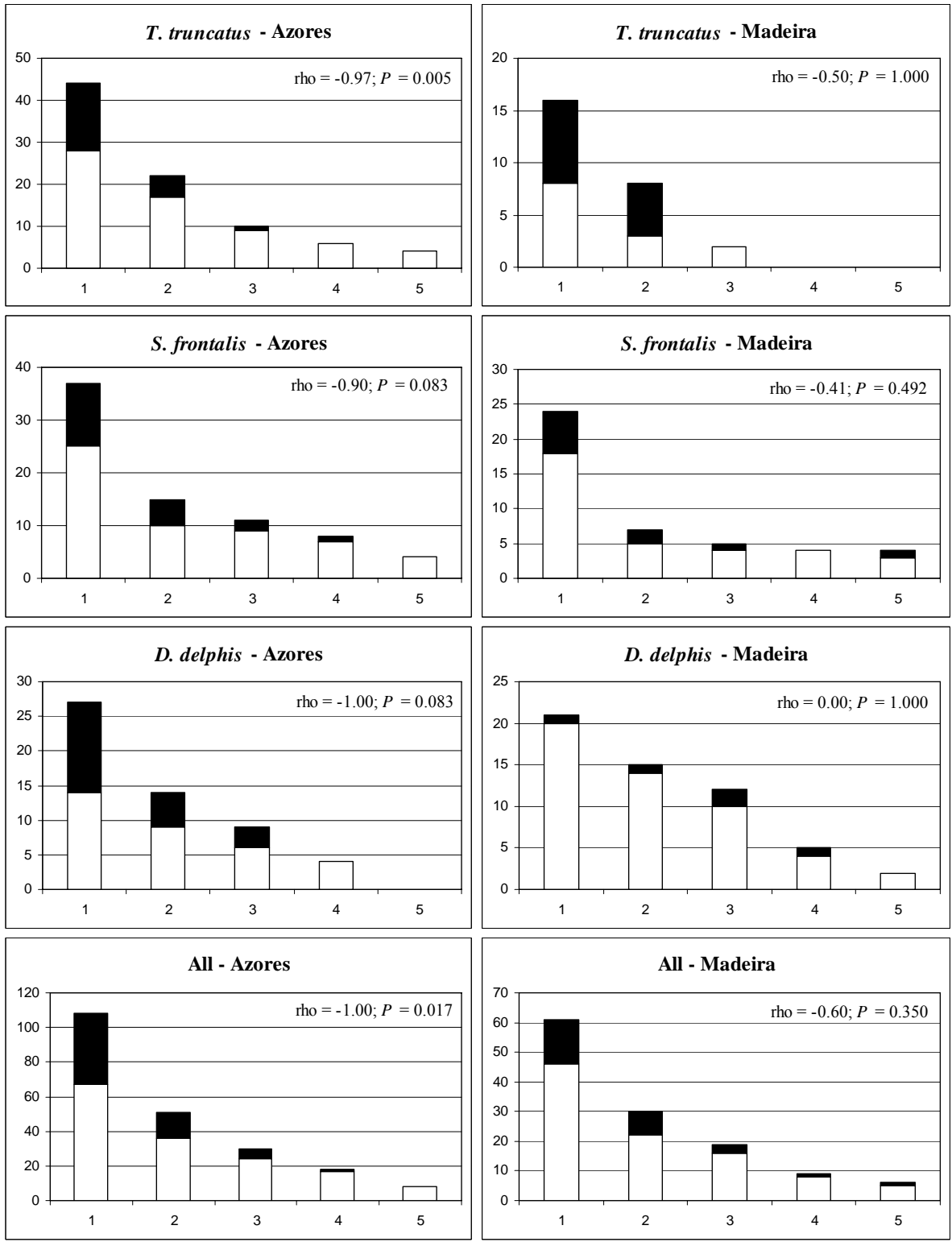